\documentclass[10pt,conference]{IEEEtran}
\usepackage{graphicx}
\usepackage[utf8]{inputenc}
\usepackage{mathtools}

\let\subparagraph\relax
\usepackage{subfiles}
\usepackage[justification=centering,font=scriptsize,skip=1pt, belowskip=1pt,aboveskip=1pt]{caption}
\usepackage{color}
\usepackage{titlesec}
\usepackage{amsmath}
\usepackage{pgfplots}
\usepackage{xspace}
\usepackage{authblk}
\usepackage{listings} 
\usepackage{xcolor}

\definecolor{codegreen}{rgb}{0,0.6,0}
\definecolor{codegray}{rgb}{0.5,0.5,0.5}
\definecolor{codepurple}{rgb}{0.58,0,0.82}
\definecolor{backcolour}{rgb}{0.95,0.95,0.92}
\lstdefinestyle{mystyle}{
    backgroundcolor=\color{backcolour},   
    commentstyle=\color{codegreen},
    keywordstyle=\color{magenta},
    numberstyle=\tiny\color{codegray},
    stringstyle=\color{codepurple},
    basicstyle=\ttfamily\footnotesize,
    breakatwhitespace=false,         
    breaklines=true,                 
    captionpos=b,                    
    keepspaces=true,                 
    showspaces=false,                
    showstringspaces=false,
    showtabs=false,                  
    tabsize=2
}

\let\OLDthebibliography\thebibliography
\renewcommand\thebibliography[1]{
  \OLDthebibliography{#1}
  \setlength{\parskip}{0pt}
  \setlength{\itemsep}{0pt plus 0.3ex}
}
\expandafter\def\expandafter\normalsize\expandafter
{%
    \normalsize
    \setlength\abovedisplayskip{-10pt}
    \setlength\belowdisplayskip{0pt}
    \setlength\abovedisplayshortskip{-10pt}
    \setlength\belowdisplayshortskip{0pt}
}

\titlespacing\section{0pt}{0pt plus 4pt minus 2pt}{0pt plus 2pt minus 2pt}
\titlespacing\subsection{0pt}{0pt plus 4pt minus 2pt}{0pt plus 2pt minus 2pt}
\titlespacing\subsubsection{0pt}{0pt plus 4pt minus 2pt}{0pt plus 2pt minus 2pt}

\begin{document}

\title{Demo: A Proof-of-Concept Implementation of Guard Secure Routing Protocol}
\author{Sanaz Taheri-Boshrooyeh \IEEEauthorrefmark{1}\IEEEauthorrefmark{3},
Ali Utkan Şahin \IEEEauthorrefmark{3}, 
Yahya Hassanzadeh-Nazarabadi \IEEEauthorrefmark{2}\IEEEauthorrefmark{3}, 
and Öznur Özkasap \IEEEauthorrefmark{3}
\\CryptoNumerics, Toronto, Canada \IEEEauthorrefmark{1}
\\DapperLabs, Vancouver, Canada \IEEEauthorrefmark{2}
\\Department of Computer Engineering, Koç University, İstanbul, Turkey \IEEEauthorrefmark{3}\\
{\{staheri14, asahin17, yhassanzadeh13, oozkasap\}}@ku.edu.tr}

\maketitle

\begin{abstract}
Skip Graphs belong to the family of Distributed Hash Table (DHT) structures that are utilized as routing overlays in various peer-to-peer applications including blockchains, cloud storage, and social networks. In a Skip Graph overlay, any misbehavior of peers during the routing of a query compromises the system functionality. \textit{Guard} is the first authenticated search mechanism for Skip Graphs, enables reliable search operation in a fully decentralized manner. In this demo paper, we present a proof-of-concept implementation of \textit{Guard} on Skip Graph nodes as well as a deployment demo scenario.
\end{abstract}

\section{Introduction}
As a DHT-based overlay, Skip Graphs \cite{aspnes2007skip} provide efficient routing functionality for peer-to-peer (P2P) systems including distributed cloud storage \cite{hassanzadeh2019decentralized}, social networks \cite{taheri2015security}, and blockchains \cite{hassanzadeh2019lightchain}. Each peer of a P2P system corresponds to a Skip Graph node with a unique identifier. In a Skip Graph with $n$ nodes, each node needs to know only $O(\log{n})$ other nodes known as neighbors. Nodes benefit from their neighbor list to locate each others (using their identifiers) in a fully decentralized manner. That is, a search query gets handed across a certain number of nodes that utilize their neighbor list to forward the query closer to the destination. The entire procedure has the message complexity of $O(\log{n})$ \cite{aspnes2007skip, hassanzadeh2016laras}. Search queries are the fundamental operation in Skip Graphs that enable the nodes to join the overlay as well as to build up complex cooperative P2P applications, e.g., P2P cloud storage.   


The \textit{malicious} nodes along a search path can jeopardize the safety of the system by conducting \textit{routing attacks} namely, by dropping, manipulating, misdirecting or falsifying a search query that they are intended to route.
Skip Graphs do not intrinsically preserve the correctness of routing in the presence of malicious actors. Besides, the applicable DHT-based solutions on Skip Graphs either degrade its decentralization \cite{Wang2006-jm,urdaneta2011survey, shabeera2012authenticated}, or increase its communication complexity \cite{Wang2012-gn}. 

To address the security of Skip Graphs against the mentioned routing attacks, we proposed \textit{Guard} \cite{boshrooyeh2017guard}, which is the first fully decentralized authenticated search mechanism for Skip Graphs. In Guard, each of the nodes involved in the routing of a search query generates and piggybacks a proof to the search message that asserts the honest behavior of that node. This enables both the search initiator as well as the nodes on the search path to verify the validity of the search result. In a \textit{Guard}-enabled Skip Graph, no malicious node can deliberately deviate from the search protocol without being detected and caught by others. 

In this demo paper, as the original contribution, we present an open-source proof-of-concept implementation of our software architecture and implementation of \textit{Guard} (in Java) \cite{guard}, and its interaction with the Skip Graph nodes it operates on. We also present the demo scenario of deploying our \textit{Guard} implementation in the production-level cloud computing environment of Amazon Web Services' Elastic Compute Cloud service (AWS EC2). 

\begin{figure}
\centering
\includegraphics[width=\linewidth]{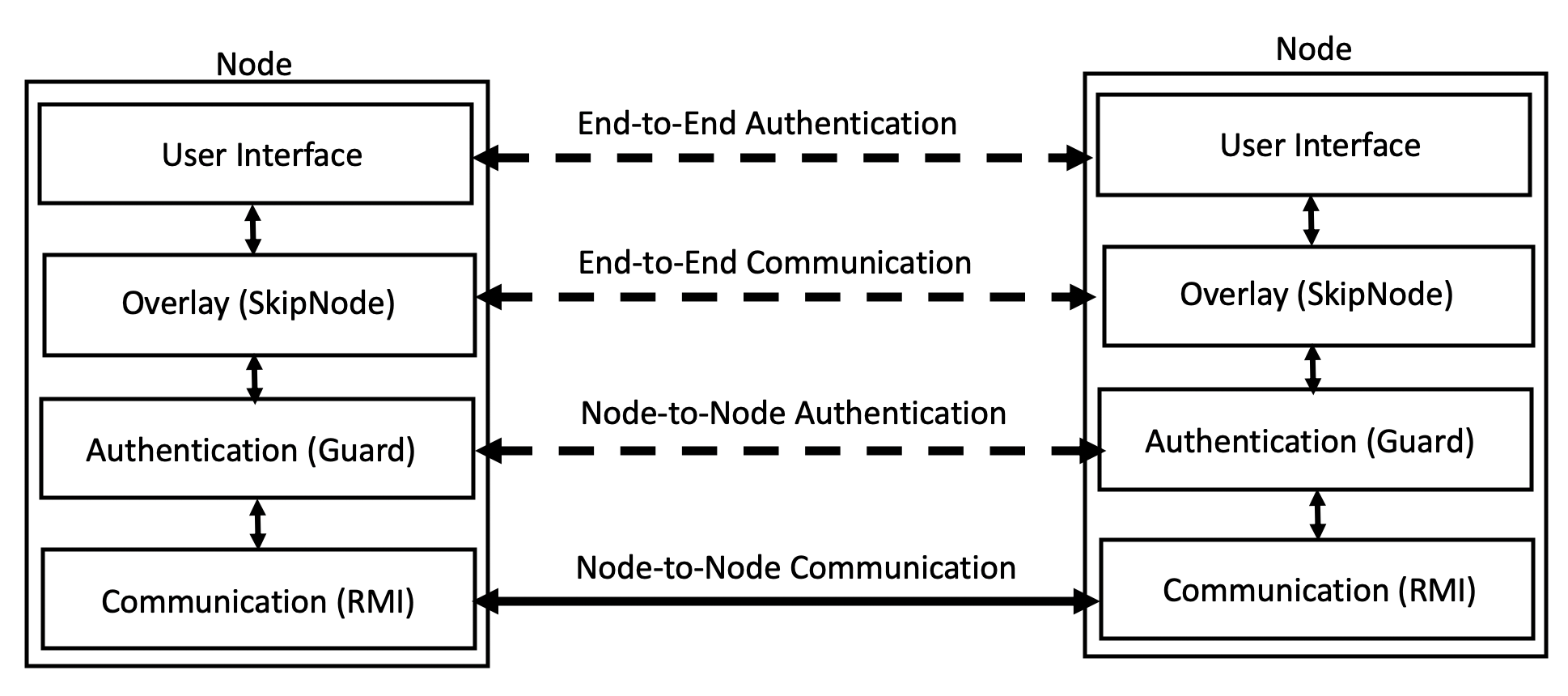}
\caption{The inter-node interactions in the layered software architecture of \textit{Guard}. The bottom-most solid horizontal arrow corresponds to a real path in the underlying network, while the other dashed horizontal arrows indicate virtual pairwise communications of the layers through the underlying network.}
\label{fig:layered-architecture}
\end{figure}

\section{Software Architecture}
The layered software architecture of our proof-of-concept implementation of \textit{Guard} is depicted in Figure \ref{fig:layered-architecture}. In this figure, the double-sided vertical arrows represent the internal interactions among the layers of a node. Similarly, the horizontal arrows show the interactions between levels of the same type on two different nodes. The layers of a node from the top to the bottom are called  \textit{User Interface}, \textit{Overlay}, \textit{Authentication}, and \textit{Communication}, respectively.

\textbf{User Interface} interacts with the user that runs the underlying peer (i.e., device) of the node. This layer receives the control commands from the user, passes them to the Overlay layer for process, and displays the obtained result from the Overlay layer. The supported operations by the User Interface layer are \textit{join} and \textit{search}. A join places the underlying peer as a Skip Graph node with a unique identifier and address in the decentralized Skip Graph overlay. A search returns the address of the node that holds the closest identifier to the search target \cite{aspnes2007skip, hassanzadeh2016laras}.

\textbf{Overlay} layer implements a Skip Graph node, and provides the User Interface with join and search operations. The join is done by receiving a unique identifier for the target node. The identifier of a peer is the hash value of its (IP) address. The search enables the overlay layer of different nodes to search for each others' identifiers in a fully decentralized manner. As the result of searching for a node's identifier, the overlay returns (IP) address of the node that holds the search result according to the search protocol \cite{aspnes2007skip, hassanzadeh2016laras}. 

\textbf{Authentication} layer is where the core operational logic of \textit{Guard} resides. It receives an overlay operation (i.e., join or search), generates authentication metadata for it and passes it to the underlying network through the Communication layer. Also, it receives the authenticated messages that are meant for the Overlay layer of the node to route or receive. Upon receiving such messages, it verifies the authentication metadata generated by the \textit{Guard} instance on the intermediate nodes, and passes the message up only if the authentication metadata is validated correctly. Structuring the software of the node in this way, both the overlay and user interface only receive messages and results that are validated against the correct execution of Skip Graph protocols on other nodes.

\textbf{Communication} provides direct node-to-node communication through the underlying network. The Communication instance of each node is uniquely identified by an (IP) address and port number. The Communication layer implements the Java Remote Method Invocation (RMI) \cite{pitt2001java} interface. This enables Communication instance of one node to directly exchange messages with the Communication instance of another node via Java RMI through the underlying network.  

\section{Sample Demo Scenario}
\begin{lstlisting}[language=Java, caption= A sample simulation schema of Guard, label=simschema]
public class Parameters {
	// search queries conducted by each node
	public static final int MESSAGE_COUNT = 1000;
	// maximum delay between consecutive queries at each node (seconds)
	public static final int WAIT_TIME = 5;
	// payload size of each message
	public static final int MESSAGE_LENGTH = 300;
	// address of simulation CONTROLLER
  public static final string CONTROLLER = <address>;
	// ...
}
\end{lstlisting}

\textbf{Configuration:} In \textit{Guard}, a simulation is configured by setting the \texttt{Parameters} class of the nodes. Listing \ref{simschema} shows a sample simulation configuration of the \textit{Guard} demo. Each node initiates one search query for a random target within the system (at most) every $5$ seconds for a total of $1000$ queries. Each query message contains $300$ bytes of payload. Note that payload of the messages is distinct than their routing information overhead (routing information are generated based on the path they are taking). To keep track of the simulation state, we define a special process known as the \texttt{CONTROLLER}, which is external to the Skip Graph overlay. There exists only a single instance of a \texttt{CONTROLLER} node for a simulation of \textit{Guard}. \texttt{CONTROLLER} acts as a centralized point of trust that synchronizes the overlay nodes with each other. \textit{Guard} is a fully decentralized protocol that operates on asynchronous communication model \cite{cristian1996synchronous} where there is no known bound over the message transmission delay among the nodes. Nevertheless, we found having a \texttt{CONTROLLER} a convenient solution only to establish large scale implementation of its proof-of-concept over cloud in a reliable manner, as well as closely tracking the simulation performance. Address of \texttt{CONTROLLER} is hardcoded in the \texttt{Parameters} class of each node. 

\textbf{Initialization:} Upon running up, each new node initiates the decentralized Skip Graph join protocol \cite{aspnes2007skip}. This protocol enables the node to interact with other nodes of the system in a fully decentralized manner and build up the overlay collaboratively. Once each node joins the overlay successfully, it registers itself to the \texttt{CONTROLLER}. Once all nodes join the system and the Skip Graph overlay is shaped completely, the \texttt{CONTROLLER} starts the initialization phase of \textit{Guard} \cite{boshrooyeh2017guard} on each node. During this phase, the Skip Graph nodes construct the authentication metadata, which is later on frequently utilized to authenticate the search queries they route. It is worth noting that the original solution of \textit{Guard} \cite{boshrooyeh2017guard} assumes no churn in the Skip Graph, i.e., once a Skip Graph node initiates the join protocol, it remains available the entire time and does not go offline or fail \cite{hassanzadeh2019interlaced}. We will address handling churn for \textit{Guard} in our future works. 

\textbf{Experiments:} Once the initialization is complete, \texttt{CONTROLLER} broadcasts an \textit{experiment request} message to every node in the system. For the sake of demonstration and to perform a fair comparison, each node performs $1000$ pairs of searches during this time upon receiving an \textit{experiment request} message. The value $1000$ corresponds to the \texttt{MESSAGE\_COUNT} field of the configuration file in Listing \ref{simschema}. In each pair of searches, the node chooses a random identifier uniformly as the search target. It then performs an unauthenticated search followed by an authenticated one. As there is no churn in the system, both searches follow the same search path. Note that \textit{Guard} has provable security in the presence of malicious colluding adversaries aiming on breaking the authenticity of search proofs \cite{boshrooyeh2017guard}. This implies that no probabilistic polynomial node on a search path can deviate from the search protocol while successfully forging a search proof that covers up for its deviation. Hence, at the level of the proof-of-concept, having the searches in pairs of authenticated and unauthenticated ones provides a fair comparison of the time overhead that \textit{Guard} applies to make the searches authenticated. To avoid congestion, however, each node waits between conducting every two consecutive pairs of searches for a period that is uniformly drawn between $1$ to $5$ (i.e., \texttt{WAIT\_TIME}). This phase is complete when every node performs the assigned $1000$ pairs of search queries and receives their results. 

\textbf{Measurements:} Once the Experiment phase is over, \texttt{Controller} collects the log file of every node and merges them into a single file. The individual log files of the nodes contain all the details of every event that has happened in their Initialization and Experiment phases. The log files are in \texttt{csv} format, which allow to make any arbitrary query and extract any parameter of interest, for example, the average query latency, the average local computation time of a node upon routing a message, and the average message size. All measurements are done in both authenticated and unauthenticated modes by filtering their corresponding logs.

\bibliographystyle{IEEEtran}
\bibliography{main}

\begin{thebibliography}{10}
\providecommand{\url}[1]{#1}
\csname url@samestyle\endcsname
\providecommand{\newblock}{\relax}
\providecommand{\bibinfo}[2]{#2}
\providecommand{\BIBentrySTDinterwordspacing}{\spaceskip=0pt\relax}
\providecommand{\BIBentryALTinterwordstretchfactor}{4}
\providecommand{\BIBentryALTinterwordspacing}{\spaceskip=\fontdimen2\font plus
\BIBentryALTinterwordstretchfactor\fontdimen3\font minus
  \fontdimen4\font\relax}
\providecommand{\BIBforeignlanguage}[2]{{%
\expandafter\ifx\csname l@#1\endcsname\relax
\typeout{** WARNING: IEEEtran.bst: No hyphenation pattern has been}%
\typeout{** loaded for the language `#1'. Using the pattern for}%
\typeout{** the default language instead.}%
\else
\language=\csname l@#1\endcsname
\fi
#2}}
\providecommand{\BIBdecl}{\relax}
\BIBdecl

\bibitem{aspnes2007skip}
J.~Aspnes and G.~Shah, ``Skip graphs,'' \emph{TALG}, 2007.

\bibitem{hassanzadeh2019decentralized}
Y.~Hassanzadeh-Nazarabadi, A.~K{\"u}p{\c{c}}{\"u}, and O.~Ozkasap,
  ``Decentralized utility-and locality-aware replication for heterogeneous
  dht-based p2p cloud storage systems,'' \emph{TPDS}, 2019.

\bibitem{taheri2015security}
S.~Taheri-Boshrooyeh, A.~K{\"u}p{\c{c}}{\"u}, and {\"O}.~{\"O}zkasap,
  ``Security and privacy of distributed online social networks,'' in \emph{IEEE
  ICDCSW, 2015}.

\bibitem{hassanzadeh2019lightchain}
Y.~Hassanzadeh-Nazarabadi, A.~K{\"u}p{\c{c}}{\"u}, and {\"O}.~{\"O}zkasap,
  ``Lightchain: A dht-based blockchain for resource constrained environments,''
  \emph{arXiv preprint arXiv:1904.00375}, 2019.

\bibitem{hassanzadeh2016laras}
Y.~Hassanzadeh-Nazarabadi, A.~K{\"u}p{\c{c}}{\"u}, and O.~Ozkasap, ``Laras:
  Locality aware replication algorithm for the skip graph,'' in \emph{IEEE NOMS
  2016}.

\bibitem{Wang2006-jm}
P.~Wang, I.~Osipkov, N.~Hopper, and Y.~Kim, ``Myrmic: Secure and robust dht
  routing,'' \emph{U. of Minnesota, Tech. Rep}, 2006.

\bibitem{urdaneta2011survey}
G.~Urdaneta, G.~Pierre, and M.~V. Steen, ``A survey of dht security
  techniques,'' \emph{ACM Computing Surveys (CSUR)}, vol.~43, no.~2, p.~8,
  2011.

\bibitem{shabeera2012authenticated}
T.~Shabeera, P.~Chandran, and S.~Kumar, ``Authenticated and persistent skip
  graph: a data structure for cloud based data-centric applications,'' in
  \emph{ACM CSS 2012}.

\bibitem{Wang2012-gn}
Q.~Wang and N.~Borisov, ``Octopus: A secure and anonymous {DHT} lookup,'' in
  \emph{{IEEE} ICDCS}, 2012.

\bibitem{boshrooyeh2017guard}
S.~T. Boshrooyeh and O.~Ozkasap, ``Guard: Secure routing in skip graph,'' in
  \emph{2017 IFIP Networking Conference (IFIP Networking) and Workshops}.\hskip
  1em plus 0.5em minus 0.4em\relax IEEE, 2017, pp. 1--2.

\bibitem{guard}
``Guard:https://github.com/staheri14/guard.''

\bibitem{pitt2001java}
E.~Pitt and K.~McNiff, \emph{Java. rmi: The remote method invocation
  guide}.\hskip 1em plus 0.5em minus 0.4em\relax Addison-Wesley Longman
  Publishing Co., Inc., 2001.

\bibitem{cristian1996synchronous}
F.~Cristian, ``Synchronous and asynchronous,'' \emph{Communications of the
  ACM}, 1996.

\bibitem{hassanzadeh2019interlaced}
Y.~Hassanzadeh-Nazarabadi, A.~K{\"u}p{\c{c}}{\"u}, and {\"O}.~{\"O}zkasap,
  ``Interlaced: Fully decentralized churn stabilization for skip graph-based
  dhts,'' \emph{arXiv preprint arXiv:1903.07289}, 2019.

\end{thebibliography}

\end{document}